\documentclass[a4paper]{jpconf}
\usepackage{graphicx}
\begin{document}
\title{A new solution for effective interaction}

\author{R Okamoto$^1$, K Suzuki$^1$, H Kumagai$^2$, S Fujii$^3$}

\address{$^1$Kyushu Institute of Technology, Kitakyushu, 804-8550 Japan}
\address{$^2$Fukuoka Institute of Technology, Fukuoka, 813-0000 Japan}
\address{$^3$Center for Nuclear Study (CNS), University of Tokyo, Wako Campus of RIKEN, Wako 351-0198, Japan
}

\ead{okamoto@mns.kyutech.ac.jp}
%---------------------------------------------------
\begin{abstract}
A new method is given for the model-space effective interaction.
Introducing a new operator in place of the $Q$-box in the Krenciglowa-Kuo (KK) 
method,  we derive a new equation for the effective interaction.
This equation can be viewed as an extension of the KK method.
We show that this equation can be solved both in iterative and non-iterative 
ways.
We observe that the iteration procedure brings about fast acceleration of 
convergence compared to 
the KK approach.
We  also find that the non-iterative calculation reproduces successfully any 
set of the true eigenvalues of the original Hamiltonian.
This non-iterative calculation can be made regardless 
of the magnitudes of the overlaps with the model space and the energy
 differences between the unperturbed energy and the eigenvalues to be solved.
\end{abstract}
%
%
%-----------------------------------------------------------------------------------------------------------
%----------------------------
\section{Introduction}
%-----------------------
In some fields of many-body physics, it is often useful to introduce an effective interaction acting in a truncated model space.
Much effort has been made both as regards formal theories and their
applications[1].

Among many approaches, we here direct our attention to the Krenciglowa-Kuo (KK)[2] and the Lee-Suzuki (LS)[3,4] methods.
These two methods are constructed by the use of the so called $Q$-box as the building block of formulation.
The KK approach has very simple structure and the solution is obtained in an iterative way.
If the iteration converges, eigenvalues are given for the states with largest overlaps with the chosen model space.
On the other hand, the LS method reproduces eigenvalues for the states which lie closest to the chosen unperturbed energy.
The LS method is complicated in structure and higher derivatives of the $Q$-box with respect to starting energy are necessary if one wants to obtain more accurate solutions.

Both of the two theories yield only certain of the true eigenvalues of the original Hamiltonian.
This restriction is not, of course, desirable.
In a formal point of view, the $Q$-box itself contains information regarding all of the true eigenvalues.
For a given model space of dimension $d$, there would be a method of reproducing any $d$ eigenvalues among all the true eigenvalues.

In this report, we want to show that a method exists for reproducing all the true eigenvalues of the original Hamiltonian if the $Q$-box is given 
accurately enough.
%-----------------------
\section{Krenciglowa-Kuo method}
%-----------------------
We assume that the Hamiltonian $H$ is composed of the unperturbed Hamiltonian $H_0$ and the perturbation $V$, i.e., $H=H_0+V$.     
The entire Hilbert space is partitioned into the $d$-dimensional model space (P space) and its complement (Q space) with the projection operators $P$ and $Q$, respectively.
We note some properties, $P+Q=1, P^2=P,Q^2=Q$ and $PQ=QP=0$.
We here assume that $H_0$ is decoupled between the P and Q spaces as 
%----------------------
\begin{equation}
\label{eq:Hamiltonian}
H_0=PH_0P+QH_0Q.
\end{equation}
%----------------------
We further assume that the P-space states have a degenerate unperturbed energy $E_0$, i.e.,
%----------------------
\begin{equation}
\label{eq:model-space-hamiltonian}
PH_0P=E_0P.
\end{equation}
%----------------------
Thus the P-space eigenvalue equation is written with the P-space effective interaction $R$ 
and the P-space eigenstate $|\phi_{k}\rangle$ as 
%----------------
\begin{eqnarray}
\label{eq:p-space-eigen}
 (E_0 P + R)|{\phi_k }\rangle  = E_k | {\phi _k } \rangle, 
 \quad \quad (k = 1,2, \cdots ,d). 
\end{eqnarray}
%------------------------
The effective interaction can be constructed by the use of $Q$-box which is an operator of energy variable $E$ and is defined as
%--------------
\begin{eqnarray}
\label{eq:Q-box}
 \hat Q(E) &\equiv& PVP + PVQ\frac{1}{{E - QHQ}}QVP.
 \end{eqnarray}
%----------------------

The Krenciglowa-Kuo (KK) approach leads to a solution for $R$ as
%-------------------
\begin{eqnarray}
 R_{\small\rm KK} &=& \sum\limits_{k = 1}^d {\hat Q} (E_k )|{\phi_k}\rangle 
 \langle {\tilde\phi_k}|\nonumber\\
\label{eq:RKK}
  &=&\sum\limits_{k = 1}^d (E_{k}-E_{0})|{\phi_k}\rangle 
 \langle {\tilde\phi_k}|,
 \end{eqnarray}
%-----------------------
where $\langle\tilde\phi_{k}|$ is the biorthogonal state of the P-space eigenstate $|\phi_k\rangle$
in Eq.~(\ref{eq:p-space-eigen}) defined through 
$\langle{\tilde\phi_k |\phi _{k'} }\rangle  = \delta_{kk'}$.
We here note that, as seen in Eq.~(\ref{eq:RKK}), the determination of the effective interaction $R_{KK}$ is equivalent to solving $d$ eigenvalues \{$E_k$\}.
In general, the K-K solution is obtained in an iterative way based on 
Eqs.~(\ref{eq:p-space-eigen}) 
and (\ref{eq:RKK}).     
It has been known that if the iteration converges, the KK solution yields $d$ eigenvalues of the states with largest P-space overlaps among the eigenstates of the original Hamiltonian $H$.

We here want to show that there is another way of obtaining the KK solution.
We consider an eigenvalue equation for a given energy $E$
%-----------------------
\begin{eqnarray}
\label{eq:Q-box-eigen}
 [E_0 + \hat Q(E)]|{\psi _k } \rangle = G_{k}(E)| {\psi_k}\rangle,
\ (k = 1,2,\cdots,d),
\end{eqnarray}
%-----------------
where \{$G_{k}(E)$\} and \{$| {\psi_k}\rangle$\} are the eigenvalues and the eigenstates, respectively.
For a given $E$, there are $d$ eigenstates because $\hat Q(E)$ is a $d$-dimensional operator.
Thus, we have $d$ functions of variable $E$, which we label in order of energy as 
$G_{1}(E) < G_{2}(E)<\cdots < G_{d}(E)$.

It may be clear from Eqs.~(\ref{eq:p-space-eigen}) and (\ref{eq:Q-box-eigen})
 that the eigenvalues 
\{$E_k$\} in Eq.~(\ref{eq:p-space-eigen}) can be given by solving equations
%-----------------------
\begin{equation}
\label{eq:EGK}
E=G_k(E), \,\,\,\,(k=1,2,\cdots ,d).
\end{equation}
%-----------------
We can prove that any solution of Eq.~(\ref{eq:EGK}) agrees with one of the true eigenvalues of $H$ and the eigenstate $| {\psi_k}\rangle$ does with $| {\phi_k}\rangle$ in Eq.~(\ref{eq:p-space-eigen}).

The solutions \{$E_k$\} to Eq.~(\ref{eq:EGK}) are obtained as crossing points of two graphs of 
$y=G_{k}(E)$ and $y=E$.
It should be noted that this procedure of solving eigenvalues \{$E_k$\} can be made independently of the  magnitude of the P-space overlap.
Therefore we can, in principle, reproduce all the true eigenvalues of $H$.
However, as we see in Eq.~(\ref{eq:Q-box}) there appear poles in $\hat Q(E)$ 
 when $E$ approaches 
one of the eigenvalues of $QHQ$.
The poles in $\hat Q(E)$ induce also  poles in $G_k(E)$.
Such a situation causes numerical instability in solving 
Eq.~(\ref{eq:EGK}) 
with $\{G_k(E)\}$ around the pole positions.
%---------------------------------------------------
\section{Extension of the Krenciglowa-Kuo method}
%--------------------------------------------------
We define an operator in terms of the $Q$-box as
%-----------------------------
\begin{eqnarray}
\label{eq:z-box}
 \hat Z(E) \equiv \frac{1}{{1 - \hat Q_1 (E)}}\left[ {\hat Q(E) - (E - E_0 ) \cdot \hat Q_1 (E)} \right],
 \end{eqnarray}
%-----------------------------
where $\hat Q_1(E)$ is the energy derivative of the $Q$-box given as
%--------------
\begin{eqnarray}
\label{eq:Q1}
 \hat Q_1(E) 
 &\equiv& \frac{d\hat Q(E)}{dE}\nonumber\\
  &=&  - PVQ\frac{1}{{(E - QHQ)^2 }}QVP.
 \end{eqnarray}
%----------------------
Hereafter we  shall refer  to $\hat Z(E)$ as the $Z$-box.
It should be noted that if $E=E_0$, the $Z$-box agrees with the first-order
 recursive solution in the LS method[4].

The $Z$-box has the following properties:
%--------------------------------------------------------------------------
\begin{enumerate}
\item 
By using Eqs.~(\ref{eq:Q-box}), (\ref{eq:RKK}) and (\ref{eq:z-box}), we have
%-------------------------------------------
\begin{eqnarray}
\label{eq:EKK-KK}
\sum\limits_{k = 1}^d {\hat Z(E_k )}|{\phi _k } \rangle 
 \langle {\tilde\phi_{k}} |
 &=&
\sum\limits_{k = 1}^d {\frac{1}{{1 - \hat Q_1 (E_k )}}
 \left[ {R_{\rm KK}  - \hat Q_1 (E_k ) \cdot R_{\rm KK}} \right]} | {\phi_k }\rangle\langle {\tilde\phi _k } | \nonumber\\
 &=& \sum\limits_{k = 1}^d {R_{\rm KK} | {\phi _k } \rangle\langle 
 {\tilde\phi _k }|}  \nonumber\\
 &=& R_{\rm KK} . 
 \end{eqnarray}
%-------------------------------------
The above fact means that, replacing $\hat Q(E)$ by $\hat Z(E)$ in 
Eq.~(\ref{eq:RKK}), a new solution for the 
effective interaction $R$ can be derived as
%-----------------------
\begin{eqnarray}
\label{eq:REKK}
 R_{\small\rm EKK} &\equiv& 
 \sum\limits_{k = 1}^d {\hat Z} (E_k )|{\phi_k}\rangle 
 \langle {\tilde\phi_k}|
 \end{eqnarray}
%-----------------------
which we call the extended Krenciglowa-Kuo (EKK) solution.

\hspace{5mm}
In a similar way as in Eq.~(\ref{eq:Q-box-eigen})
we  solve the $d$-dimensional P-space eigenvalue equation for
the $Z$-box for a given variable $E$ as 
%-------------------------------- 
\begin{eqnarray} 
\label{eq-Z-eigenvalue}
 \left\{ {E_0  + \hat Z(E)} \right\}| {\chi_k }\rangle=F_{k}(E)
|{\chi_k}\rangle,\ (k = 1,2,\cdots,d) , 
 \end{eqnarray}
%---------------------------
where $\{F_{k}(E)\}$ are the eigenvalues and $\{|{\chi_k}\rangle\}$ are the eigenstates.
We here label {$F_k(E)$} in order of  energy as
 $F_{1}(E) < F_{2}(E)<\cdots < F_{d}(E)$.
In the same way as in Eq.~(\ref{eq:EGK}), the true eigenvalues \{$E_k$\} are given by solving the equations
%-----------------------
\begin{equation}
\label{eq:EFK}
E=F_k(E), \,\,\,\,(k=1,2,\cdots ,d).
\end{equation}
%
%
%-----------------
\item 
 We can prove that if $E$ is one of the true eigenvalues \{$E_k$\} 
 satisfying
 Eq.~(\ref{eq:RKK}),
the energy derivative of $\hat Z(E)$ becomes zero, and we have
%-------------------------
\begin{eqnarray}
\label{eq:Fk-derivative}
\quad \frac{{dF_k (E)}}{{dE}} = 0.
\end{eqnarray}
%--------------
This property of the $Z$-box plays important roles in two ways:

(a)\,\, The two equations (\ref{eq:Q-box-eigen}) and (\ref{eq-Z-eigenvalue}) are not identical.
The operator $\hat Z(E)$ is different from $\hat Q(E)$ as a P-space operator. 
Therefore, Eq.~(\ref{eq:EFK}) based on the $Z$-box has possibly excessive solutions, or spurious solutions, different from the true eigenvalues \{$E_k$\}.
However, we can remove easily these spurious solutions according to the condition in Eq.~(\ref{eq:Fk-derivative}).

(b)\,\,In the same way as in the KK approach we can derive the effective interaction $R_{\rm EKK}$ in an iterative way.
The iterative calculation in this approach can be made only by replacing $\hat Q(E)$ in the KK method by $\hat Z(E)$.
It must be pointed out that, due to the property of  Eq.~(\ref{eq:Fk-derivative}), the iteration based on $\hat Z(E)$ converges quite rapidly as we see in Table 1.
The new method can be understood as an application  of the Newton-Raphson (NR) method which is  used widely for accelerating convergence in solving  non-linear 
equations iteratively. 
\item 
In the vicinity of the poles in $\hat Q(E)$, $\hat{Q}_{1}(E)$ has  the dominant contribution in $\hat Z(E)$, which may be clear from Eqs.~(\ref{eq:Q-box}), (\ref{eq:z-box}) and 
(\ref{eq:Q1}). 
Resultantly $\hat{Z}(E)$ becomes
%---------------
\begin{eqnarray}
 \hat{Z}(E) &\approx& (E - E_0 )P,
 \end{eqnarray}
%--------------
and thus there appear no poles in $\hat Z(E)$.
Therefore, the functions \{$F_k(E)$\} are continuous and differentiable for 
any $E$. This fact of the new approach guarantees stability in numerically solving the crossing points
 of two graphs of $y=F_k(E)$ and $y=E$.
We want to emphasize here that the procedure of solving \{$E_k$\} is regardless of the properties of the eigenstates of  the original Hamiltonian $H$, i.e., the P-space overlap or the position in energy.
\end{enumerate}
%---------------------------------------
%--------------------

%\input{okamoto1.tex}
\begin{figure}
\begin{center}
%\begin{minipage}{17pc}
%\includegraphics[width=17pc]{Gk0.2.eps}
\includegraphics[width=17pc]{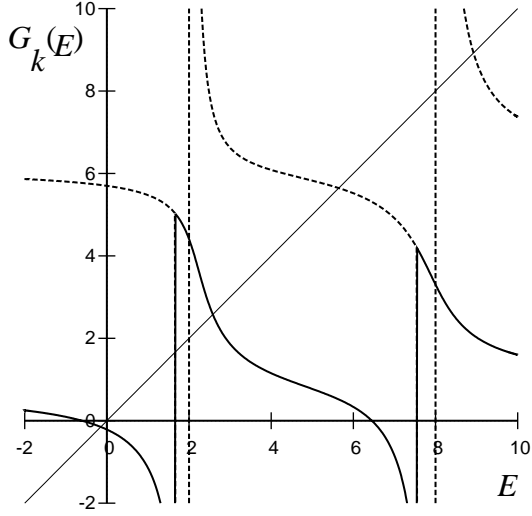}
%\input{okamoto1.tex}
%\end{minipage}
\end{center}
%\begin{figure}
\vskip -2pt 
\caption{%\vspace{-2cm}
\label{fig-1}Dependence of $G_{k}(E)$ on $E$ with $x=0.2$.
The graphs of $ y=G_{1}(E)$ and $y=G_{2}(E)$ are shown in solid and broken lines, respectively. The direct line denotes the graph of $y=E$.}
\end{figure}

%------------------
%--------------------
\begin{figure}
\begin{center}
\includegraphics[width=17pc]{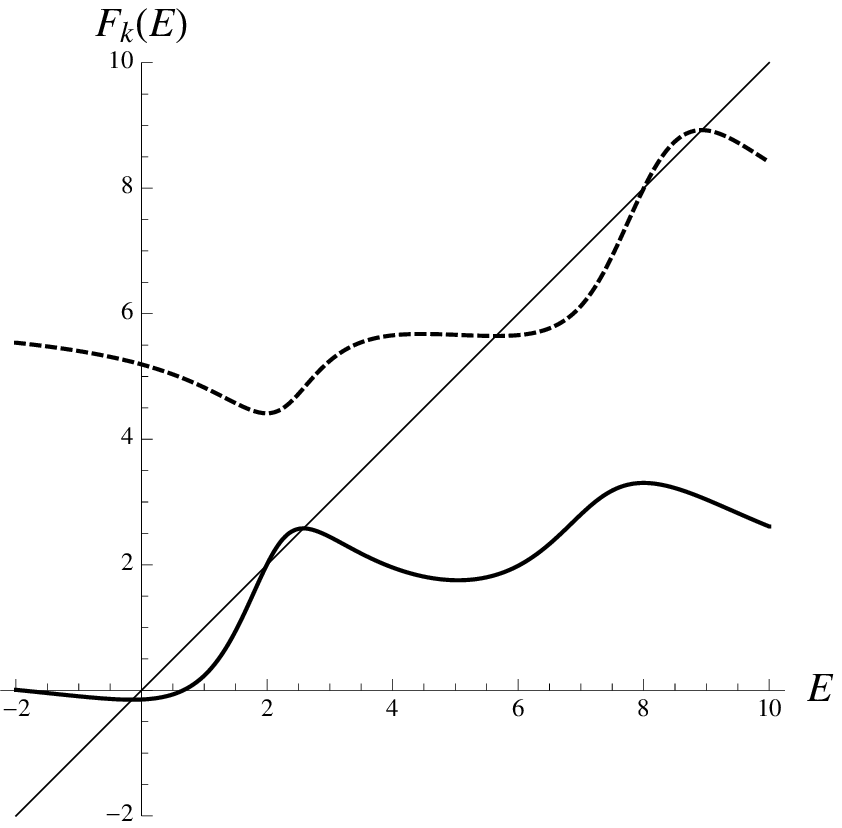}
\end{center}
\caption{\label{fig-2}Dependence of $F_{k}(E)$ on $E$ with $x=0.2$.
The graphs of $ y=F_{1}(E)$ and $y=F_{2}(E)$ are shown in solid and broken lines, 
respectively. The direct line denotes the graph of $y=E$.}
\end{figure}
%----------
%--------------------------------------
%%----------------------------
\begin{table}[h]
\caption{\label{table-1} Convergence of the eigenvalue of the lowest-lying state obtained with the KK and EKK methods for the model Hamiltonian with $x=0.05.$ Correct digits with the KK and EKK methods are given in $n$-th iteration step. The starting energies for $(E_{1}, E_{2})$ are taken to be $(0.0, 0.0)$. The notation c indicates convergence to more than fifteen decimal places. The exact eigenvalue here is 0.8904504858869942.}
\begin{center}
\begin{tabular}{clll}
\br
$n$ &correct digits (KK)&correct digits ({\rm EKK}) \\
\mr
1  & 0.9$\cdots$            & 0.89$\cdots$    \\
2  & 0.89$\cdots$          & 0.890450$\cdots$   \\
3  & 0.8904$\cdots$         & 0.89045048588699$\cdots$\\
4  & 0.890450$\cdots$       & \hspace{5mm}  c  \\
5  & 0.8904504$\cdots$     & \hspace{5mm}  c   \\
6  & 0.89045048$\cdots$    & \hspace{5mm}  c   \\
7  & 0.8904504858$\cdots$  & \hspace{5mm}  c   \\
\br
\end{tabular}
\end{center}
\end{table}
%-----------------------------
%-----------------------------------------------------------
\section{A schematic model analysis}
%-----------------------------------------------------------
In order to obtain some assessments of the new approach we consider a model
problem for which exact solutions can be derived easily. The model Hamiltonian we adopt here is given with the coupling strength $x$, the dimension of the entire space $n=4$, the dimension of the P space $d=2$ and the unperturbed energy $E_0=1$ as
%---------------------------
\begin{eqnarray}
PH_{0}P 
  &=&\left( 
      {\begin{array}{*{20}c}
       1 & 0  \\
       0 & 1  \\
       \end{array}} 
     \right),\ 
PVP 
  = \left( 
    {\begin{array}{*{20}c}
       0  & {5x}  \\
     {5x} & {25x}  \\
     \end{array}} \right),\;
PVQ 
  =\left( 
     {\begin{array}{*{20}c}
      { - 5x} & {5x}     \\
      {5x}    & { - 8x}  \\
     \end{array}} 
    \right),
\end{eqnarray}
and
\begin{eqnarray}
QVP
   &=&\left( 
      {\begin{array}{*{20}c}
       {- 5x} & {5x}    \\
       {5x}   & { - 8x} \\
      \end{array}} 
    \right),\  
 QHQ 
  = \left( 
     {\begin{array}{*{20}c}
       {3 - 5x} & x        \\
          x     & {9 - 5x} \\
      \end{array}} 
    \right). 
\end{eqnarray}
This model Hamiltonian $H$ was introduced many years ago by Hoffmann et al.[5] and the structure of $H$ was investigated in Ref.~[6].
We first depict,  in Figs. 1 and 2, the dependence of the functions $G_k(E)$ and $F_k(E)$, respectively, on the energy variable $E$ with the strength $x=0.2$.
One may see in Fig. 1 that there are two poles in the graph of $G_k(E)$.
On the other hand, poles disappear in the graph of $F_k(E)$ associated with the $Z$-box.
One observes four crossing points between $y=G_k(E)$ and $y=E$ except for pole positions,
 and six crossing points between $y=F_k(E)$ and $y=E$ for $k=1,2$.
The spurious solutions in Fig. 2 for $y=F_k(E)$ can be easily removed by the 
condition that the energy derivative at the crossing points should be zero.

Next we show in Table 1 the results in the iterative method for the KK and EKK solutions.
The convergence rate in the EKK method is much higher than that in the KK method.
The EKK method reaches the convergence to fourteen decimal places after only three
 iterations.
However, the iterative calculation is not always recommendatory, because convergence conditions are not always clear and we cannot control these conditions.
In this respect, it would be recommended to solve directly the crossing points 
of $y=G_k(E)$ or $y=F_k(E)$ and $y=E$.
We could reproduce all the true eigenvalues of $H$ by calculating the crossing 
points in a non-iterative way, for example, the bisection method.
%
%
%-----------------------------------------------------------
\section{Conclusion}
%-----------------------------------------------------------
In many cases of calculating the effective interaction, iteration and/or recursion methods, for example, the KK and LS methods, have been employed.
However, one has seen that these methods reproduce only certain of the exact solutions.
In general, it is quite difficult to control convergence of the iteration and/or the recursion.
Moreover, these methods do not work well if a solution to be solved lies in the vicinity of the pole position.
Such a case takes place actually in the calculation using the $Q$-box as in the KK approach.

In the present report we have proposed a new approach by introducing a new
 operator, the $Z$-box, and the functions $\{F_k(E)\}$.
Characteristics of this approach are as follows: 
(a)\,\, The eigenvalues in the model space can be obtained by calculating 
the crossing points of two graphs $y=F_k(E)$ and $y=E$.
The functions $\{F_k(E)\}$ give us all the information on the eigenvalues of the original Hamiltonian $H$.
The crossing points can be calculated by using a non-iterative method, such as the bisection method.
In this case convergence problem does not occur anymore.
(b)\,\,The solutions can  be also calculated by applying an iteration method.
In this case, the use of $F_k(E)$ accelerates  convergence compared to the usual KK approach.
We want to notice that this approach is understood essentially as an application of the Newton-Raphson method for solving a non-linear equation.
(c)\,\,The function $F_k(E)$ is continuous and differentiable for any of $E$.
Instability or difficulty due to the presence of poles does not occur anymore as long as we use the function $F_k(E)$.

We may conclude that the present formulation on the procedure for obtaining the effective interaction removes some difficulties encountered in the usual approaches, such as the KK and LS methods.
This approach makes us free from divergence of iteration or the pole problem, and therefore it may have possibility of applying widely to many-body problems. 
%
%
%
%-------------------------------------------------
\section{References}
%-------------------------------------------------

\smallskip
%%%%%%%%%%%%%%%%%%%%%%%%%%%%%%%%%%
\end{document}